\documentclass{svmult}

\usepackage{makeidx}         
\usepackage{graphicx}        
\usepackage{multicol}        
\usepackage[bottom]{footmisc}

\makeindex             

\usepackage[utf8]{inputenc}
\usepackage[english]{babel}

\usepackage{cite}
\usepackage{url}

\usepackage{amsmath}
\usepackage{amsfonts}
\usepackage{amssymb}

\usepackage{color}

\begin{document}

\title*{The stability of a graph partition: \\
A dynamics-based framework for community detection}
\titlerunning{The stability of a graph partition}

\author{Jean-Charles Delvenne \inst{1,2} \and
Michael T. Schaub\inst{3,4}\and
Sophia N. Yaliraki \inst{3} \and
Mauricio Barahona \inst{4}}
\authorrunning{Delvenne et al}
\institute{Institute of Information and Communication Technologies, Electronics and Applied Mathematics (ICTEAM) and Center for Operations Research and Optimisation (CORE),
Universit\'{e} catholique de Louvain, Louvain-la-Neuve, Belgium
\and Namur Center for Complex Systems (naXys), Facult\'{e}s Universitaires Notre-Dame de la Paix, Namur, Belgium
\and Department of Chemistry, Imperial College London, London, United Kingdom
\and Department of Mathematics, Imperial College London, London, United Kingdom
\\
\texttt{jean-charles.delvenne@uclouvain.be},
\texttt{michael.schaub09@imperial.ac.uk},
\texttt{s.yaliraki@imperial.ac.uk},
\texttt{m.barahona@imperial.ac.uk}}
%
%
\maketitle

\section{Introduction}

Recent years have seen a surge of interest in the analysis of complex systems.
This trend has been facilitated by the availability of relational data and the increasingly powerful computational resources that can be employed for their analysis.
A unifying concept in the study of complex systems is their formalisation as \textit{networks} comprising a large number of non-trivially interacting agents.
By considering a network perspective, it is hoped to gain a deepened understanding of system-level properties beyond what could be achieved by focussing solely on the constituent units.
Naturally, the study of real-world systems leads to highly complex networks and a current challenge is to extract intelligible, simplified descriptions from the network in terms of relevant subgraphs (or \textit{communities}), which can provide insight into the structure and function of the overall system.

Sparked by seminal work by Newman and Girvan~\cite{Newman2004,Newman2006}, an interesting line of research has been devoted to investigating modular community structure in networks, revitalising the classic problem of graph partitioning.
In a prescient piece, Simon \cite{Simon1962} hypothesised that a modular structure exhibits evolutionary advantages.
In particular, the ability to reuse components or to facilitate efficient local processing have been extensively studied thereafter.
In recent years, a myriad of studies have gone on to detect communities and hierarchies in real world systems, ranging from social systems to technological and bio-chemical systems (for a recent review see~\cite{Fortunato2010}).
Apart from the interest in the structure of these networks, the hope is that by finding a meaningful decomposition of a network we will gain understanding of the relationship between the structure and the functional (dynamical) behaviour of the system.
As Francis Crick put it: ``If you want to understand function, study structure.''

However, modular or community structure in networks has notoriously evaded rigorous definition.
The most accepted notion of community is perhaps that of a group of elements which exhibit a stronger level of interaction within themselves than with the elements outside the community.
This concept has resulted in a plethora of computational methods and heuristics for community detection.
Nevertheless a firmly grounded theoretical understanding of most of these methods, in terms of how they operate and what they are supposed to detect, is still lacking to date.
For a meaningful application of community detection methods, such an understanding is however essential.
In the following, we will develop a dynamical perspective towards community detection enabling us to define a measure named the \textit{stability of a graph partition}~\cite{Delvenne2010,Lambiotte2009}.
It will be shown that a number of previously ad-hoc defined heuristics for community detection can be seen as particular cases of our method providing us with a dynamic reinterpretation of those measures.
Our dynamics-based approach thus serves as a unifying framework to gain a deeper understanding of different aspects and problems associated with community detection and allows us to propose new dynamically-inspired criteria for community structure. Further discussion on dynamics-based community detection, along with a description of some earlier methods, is proposed elsewhere in this book~\cite{lambiottechapter}. 

The chapter is organised as follows.
The stability of a graph partition is defined and analysed in Section 2.
In Section 3, we show how it encompasses several other community detection methods.
Section 4 discusses how to find the optimal partition with respect to stability and how to evaluate the meaningfulness of the results.
Section 5  illustrates different capabilities of the method for multiscale community detection through applications to diverse examples.

\subsubsection{Notation}
For simplicity, in the following we consider only undirected, connected and non-bipartite graphs with $N$ nodes and $E$ edges.
It is important to remark that the methodology extends seamlessly to directed graphs as discussed in \cite{Lambiotte2009}.
We use the following notation hereafter.
The topology of a graph is encoded in the weighted adjacency matrix $A\in \mathbb R^{N\times N}$, where the weight of the link between node $i$ and node $j$ is given by $A_{ij}$ and $A= A^T$ due to the undirected nature of the graph.
The weighted degrees (or strengths) of the nodes are given by the vector $d = \ A \mathbf{1}$, where $\mathbf{1}$ is the $N \times 1$ vector of ones.
We further define the diagonal matrix $D = \text{diag}(d)$ and denote the total weight of the edges by $m = d^T\mathbf{1} / 2$.
Finally, the \textit{combinatorial} graph Laplacian is defined as $L=D-A$, while the \textit{normalised} graph Laplacian is defined as $\mathcal{L}=D^{-1/2} L D^{-1/2}$.
Both Laplacians are symmetric nonnegative definite, with a simple zero eigenvalue when the graph is connected~\cite{chung1997spectral}.

\section{Dynamics on graphs and community detection: the stability of a graph partition}

As outlined above, a desirable outcome of community detection is to find a meaningful, simplified structural decomposition of a network that can shed light on the functional (dynamical) behaviour of the network and its components.
Conversely, when a dynamics takes place on a network, it will be constrained by its graph structure and could potentially reveal features of the structural organisation of the network.
This notion, i.e., that structure and dynamical behaviour are inextricably linked and influence each other, is the seed from which our perspective on community has emerged.

The central idea underpinning our dynamical approach is the following: Considering a diffusion process on a network, we ask ourselves what this simplest form of dynamics can reveal about the underlying community structure of the graph.
This idea is readily illustrated by the example of a vessel filled with water in which one were to put a small droplet of ink and observe how it diffuses.
If the container has no structure, the dye would diffuse isotropically.
If the container is structured (e.g., compartmentalised or comprising a set of smaller vessels connected via some tubing), the dye would not spread isotropically but would rather get transiently trapped in certain regions for longer times until it eventually becomes evenly distributed throughout the whole vessel.
Therefore by observing the time dynamics of this diffusion process we can gain valuable information about the structural organisation of the container.

\subsection{Defining the measure: stability as an autocovariance}
Let us now make more precise some of these ideas and consider a Markovian diffusion process defined on the graph.
One of the advantages of the method is that it allows us to choose the dynamics used to reveal the structure of the network from a variety of discrete or continuous time processes (see an extended discussion of these issues in Refs.~\cite{Lambiotte2009,Lambiotte2011}).
The dynamics chosen will have distinct dominant diffusion paths and thus will lead to the the emergence of potentially different community structure.
The flexibility in the choice of dynamics allows us to incorporate any \textit{a priori} information we might have about the intrinsic dynamics in the system under study or to base our detection on a particular heuristic for the defining characteristics of the communities.
In this work, we discuss only the generic case of an \textit{unbiased} random walk governed by the following dynamics:
\begin{equation}
 \mathbf{\dot p} = - \mathbf{p}\;[D^{-1}L]
\end{equation}
in continuous time or, alternatively, by
\begin{equation}
\mathbf{p}_{t+1} =\mathbf{p}_tD^{-1}A \equiv \mathbf{p}_t M
\end{equation}
for the case of discrete time.
Here $\mathbf{p}$ denotes the $1\times N$ dimensional probability vector and under the assumptions made above (connected, undirected, and non-bipartite graph) both dynamics converge to a unique stationary distribution given by $\pi = d^T/2m$.
While the discrete-time random walker jumps from one node to the next at unit-time intervals, the continuous-time random walker waiting time at a node before the next jump is a continuous memoryless random variable distributed exponentially with unit expectation.
Whenever a jump occurs, the jump probabilities between nodes are identical for both processes.
Hence the discrete-time random walk can also be interpreted as an approximation of its continuous-time counterpart (see also Section \ref{sec:free_energy}) and our numerics show that the two processes lead to similar community structures in most graphs of interest.

Consider also a given (hard) partition of this graph into $c$ communities.
This partition is encoded in a $N \times  c$ indicator matrix $H$ with $H_{ij} \in \{0, 1\}$, where a 1 denotes that node $i$ belongs to community $j$.
Observe that, for any $N \times N$ matrix $B$, $[H^T B H]$ is the $c \times c$ matrix whose $k \, l$--th entry is the sum of all entries $B_{ij}$ such that node $i$ is in community $k$ and node $j$ is in community $l$. It is also easy to see that $H^T H = \text{diag} (n_k)$, where $n_k$ is the number of nodes in community $k$.

Now we define the \textit{clustered autocovariance} matrix of the diffusion process at time $t$ as:
\begin{equation}
\label{eq:autocovariance}
  R(t) = H^T\left[\Pi P(t)-\pi^T\pi\right]H ,
\end{equation}
where $\Pi = \text{diag}(\pi)$
and $P(t)$ is the $t$-step transition matrix of  the process: $P(t) = \exp(-tD^{-1}L)$ for continuous-time and $P(t) = M^t$ for discrete-time.

To see that $R(t)$ is indeed an autocovariance matrix associated with the $c$ communities of a partition, consider the $N$-dimensional vector $X(t)$, a random indicator vector describing the presence of a particle diffusing under the above dynamics, i.e. $X_k(t) =1$ if the particle is in node $k$ at time $t$ and zero otherwise.
Given a partition $H$, $Y(t) = H^TX(t)$ is the community indicator vector. Using the definition of the transition matrix, it follows that:  
\begin{equation}
R(t) 
=H^T\text{cov}\left[ X(\tau),X(\tau+t)\right]H  = \text{cov}\left[ Y(\tau),Y(\tau + t)\right],
\end{equation}
where we have made use of the linearity of the covariance.

If the graph has well defined communities \textit{over a given time scale} we expect that the state is more likely to remain within the starting community with a significant probability over that time scale. Therefore the value of $Y_i(0)$ is positively correlated with the value of $Y_i(t)$ for $t$ in that time scale leading to a large diagonal element of $R(t)$, and hence a large trace of $R(t)$.
In other words, if we observe the Markov process only in terms of a colouring assigned to each community, a good partition is indicated by a slow decay of the autocovariance of the observed signal.
This statement presents a dual (but equivalent) view to the description of a community in terms of a subgraph where the diffusion probability gets transiently trapped.

We now define the \textit{stability of a partition} encoded by $H$ at time $t$ as:
\begin{equation}
\label{eq:stability}
 r(t, H) =  \text{trace } R(t),
\end{equation}
which is a global quality function for a given graph and partition that changes as a function of time (according to the chosen dynamics).
In view of the formalism outlined above, stability serves as a quality function to evaluate the goodness of a partition in terms of the persistence of the probability flows in the graph at time $t$.

A brief technical aside. The definition of stability in the discrete-time case
\begin{equation}
\label{eq:stability_disc}
 r_t (H) = \min_{0\leq s\leq t} \text{trace } \left[ H^T\left[\Pi M^s-\pi^T\pi\right]H \right] 
\end{equation}
includes a minimisation over the time interval $[0,t]$, which is not necessary in the continuous-time case.\footnote{In the continuous case, $\text{trace } R(t)$ is monotonically decreasing with time. To prove this, note that $P(t)= D^{-1/2} \exp(-t \mathcal{L}) D^{1/2}$ and $d (\text{trace } R(t))/dt = - \text{trace } H^T\Pi D^{-1/2} \mathcal{L}^{1/2} \exp(-t \mathcal{L}) \mathcal{L}^{1/2} D^{1/2}H = - \text{trace }  H^T  D^{1/2} \mathcal{L}^{1/2} \exp(-t \mathcal{L}) \mathcal{L}^{1/2} D^{1/2}H /2m$. This is obviously strictly negative since the matrix $\exp(-\mathcal{L})$ is symmetric positive definite.}
This technicality ensures maximum generality of the definition but in most cases of interest it is not necessary\footnote{In particular cases, such as a bipartite graph, $\text{trace } R_s$ can oscillate in the discrete-time case, indicating poor communities or even `anti-communities' with a rapid alternance of random walkers between communities. We therefore take the lowest point of the $R_s$ over the interval as the quality function.}. Indeed, numerical experiments~\cite{LeMartelot2012} have shown that  $$r_t (H) = \min_{0\leq s\leq t}\text{trace } R_s \approx \text{trace } R_t$$ for the discrete-time case too. In the rest of this chapter, we assume this to be valid.  

The stability $r(t,H)$ can be used to rank partitions of a given graph at different time scales, with higher stability indicating a better partition into communities at a specific time scale. Alternatively,  $r(t,H)$ can be maximised for every time $t$ in the space of all possible partitions---resulting in a sequence of partitions, each of which is optimal over some time interval.
Although, as is the case for many graph theoretical problems, this optimisation is NP-hard, a variety of optimisation heuristics for graph clustering can be used (see Section \ref{sec:optimisation}).
The effect of time in the stability measure is intuitive: as time increases, the Markov process explores larger regions of the graph, such that the Markov time acts as a resolution parameter that enables us to identify community structure at different scales.
In general, the relevant partitions become coarser as the Markov time increases.
Importantly, stability does not aim to find \textit{the} best partition for the graph, but rather tries to reveal relevant clusterings at different scales through the systematic zooming process induced naturally by the dynamics.
A relevant partition should be both persistent over a comparably long time scale and robust with respect to slight variations in the graph structure and/or the optimisation~\cite{Lambiotte2010,Delmotte2011} (see also Section \ref{sec:optimisation}).
We remark that the systematic sweeping across scales provided by the Markov dynamics is a fundamental ingredient in our approach leading to a multi-scale analysis of the community structure.

\subsection{Stability from a random walk perspective}

Stability can be readily interpreted in terms of a random walk by noting that each entry $[R(t)]_{ij}$ of the clustered autocovariance~\eqref{eq:autocovariance} is the difference between two quantities. The first is the probability for a random walker to start in community $i$ at stationarity and end up in community $j$ after time $t$, while the second is the probability of two independent walkers to be in $i$ and $j$ at stationarity.
In this view, communities correspond to groups of nodes within which the probability distribution of the Markov process is more contained after a time $t$ than otherwise expected at stationarity.
To make this perspective more explicit, one can rewrite the stability of a partition $H$ with communities $\mathcal{C}$ as:
\begin{equation}
r(t,H) = \sum_{\mathcal C=1}^c P_\text{AtHome}(\mathcal C,t)-P_\text{AtHome}(\mathcal C, \infty),
\end{equation}
where $P_\text{AtHome}(\mathcal{C},t)$ is the probability for a random walker to be in the same community $\mathcal{C}$ at time zero and at time $t$ (possibly leaving and coming back a number of times in-between).
By ergodicity, the walker at infinite time holds no memory of its initial position, and $P_\text{AtHome}(\mathcal C, \infty)$ coincides with the probability of two independent walkers to be in the same community.

\subsection{A free energy perspective: stability as a trade-off between entropy and generalised cut} \label{sec:thirdpers}
Stability may also be interpreted as a trade-off between an entropy and a generalised cut measure, thus providing a link with the concept of free energy.
To consider this perspective, we first observe that stability at time $t=0$ is essentially equivalent to the so called Gini-Simpson \textit{diversity index}\cite{simpson1949measurement}:
\begin{equation}
 r(0,H) = \text{trace}\left[H^T \left( \Pi -\pi^T\pi \right) H \right] = 1-\sum_{\mathcal{C}=1}^c (\pi \vec h_\mathcal{C})^2 \equiv \text{Diversity},
\end{equation}
where $\vec h_\mathcal{C}$ is the $\mathcal{C}$-th column of the indicator matrix $H$.
The diversity index (or equivalent quantities up to some simple transformation) has appeared under numerous names in various fields: the Hirschman-Herfindahl index in economics~\cite{hirschman1964paternity}, the R\'{e}nyi entropy of order 2 in information theory~\cite{renyi1961measures}, or the Tsallis entropy of parameter 2  in non-extensive thermodynamics~\cite{tsallis1988possible}.
This quantity may be interpreted as a measure of entropy and favours partitions into many communities of equal sizes (in terms of degree weights for the unbiased dynamics) over partitions where one community is much bigger (in terms of degree weights) than the others.
For instance, the diversity index of a $k$-way equal-size partition is $1-(1/k)$, while a partition with a very dominant community and $k-1$ small communities has a diversity index of almost zero.
The diversity index is minimal for the all-in-one partition and maximal for the partition into one-node communities.

Secondly, consider the variation of stability between time zero and time $t$:
\begin{equation}
r(0,H)- r(t,H) = 1 - \text{trace}\left[H^T\Pi P(t)H\right] \equiv \text{Generalised Cut}(t).
\end{equation}
This quantity is the fraction of walks of length $t$ that start and end in two different communities and may be thought of as a generalised cut size, i.e., the fraction of edge weights hanging between communities for a graph with adjacency matrix $\tilde A=\Pi P(t)$ induced by the dynamics of the diffusion process at time $t$~\cite{Lambiotte2009, Lambiotte2011}.

It then follows that stability can be written as
\begin{equation}
 r(t,H) = \text{Diversity} - \text{Generalised Cut}(t),
\end{equation}
which provides us with an interpretation of stability as a trade-off between an entropy and a cut measure moderated by the Markov time.
We will see below that when considering the linearised version of the stability $r(t,H)$, this interpretation allows us to write stability as a kind of free energy that balances an entropy (diversity) and an energy (cut) with the time $t$ playing the role of an inverse temperature.

\section{Stability as a unifying framework for other community detection methods}\label{sec:relations}

Stability provides an unifying framework for a number of different graph partitioning and community detection techniques and heuristics that have been postulated in the literature under different premises.
We now highlight some of these relations to provide further insight into the stability framework, while at the same time establishing a dynamical reinterpretation of some key measures widely used in community detection.
Most of the results shown below can be found in Refs.~\cite{Lambiotte2009,Delvenne2010}.

In the following, we will use shorthand to distinguish between the discrete-time version of stability (denoted as $r_t(H)$) and the continuous-time version (denoted by $r(t,H)$).

\subsection{Discrete-time stability at time one is modularity}
Modularity~\cite{Newman2004} is usually defined as:
\begin{equation}
\text{Modularity}= \dfrac{1}{2m} \sum_{\mathcal{C}} \sum_{i,j \in \mathcal{C}} A_{ij} - \dfrac{k_i k_j}{2m}
\label{eq:modu}
\end{equation}
where the sum is taken over nodes $i,j$ that are in the same community, $k_i$ is the degree of node $i$ and $m$ is the number of edges. Modularity is  a popular cost function which is maximised to find the best partition of a graph.
Noting that the stationary probability $\pi_i = k_i/2m$, it follows easily from our formalism that modularity is equivalent to discrete-time stability at time $1$:
\begin{equation}
r_1(H) = \text{trace }\left[ H^T\left(\dfrac{A}{2m} -\pi^T\pi\right)H\right] = \text{Modularity}.
\end{equation}

Furthermore,  from Section~\ref{sec:thirdpers}, it is easy to see that Generalised Cut at time $t=1$ in the discrete-time case is simply the fraction of edges between communities, i.e., the (standard) Cut:
\begin{equation}
\label{eq:gen_cut}
\text{Generalised Cut}(1) =  1 - \text{trace}\left[H^T\frac{A}{2m}H\right]= \text{trace}\left[H^T \frac{L}{2m}H\right ] =  \text{Cut},
\end{equation}
whence it follows that
\begin{equation} \label{Eq:rDT1}
 r_1(H) = \text{Diversity} - \text{Cut} = \text{Modularity}.
\end{equation}
Therefore, Modularity can be seen as a compound quality function with two competing objectives: minimize the Cut size while at the same time try to maximise the Diversity Index (which favours a large number of equally-sized communities) thus resulting in more balanced partitions.

\subsection{Stability at large time is optimised by the normalised Fiedler partition}

The asymptotic behaviour of stability at large $t$ is determined by the spectral properties of the graph and leads to the dominance of the Fiedler bipartition (a classic heuristic in graph partitioning) as $t \to \infty$.
Consider first the discrete-time case.
Elementary spectral theory implies that the eigenvalues $\lambda_i$ of $M \equiv D^{-1}A$ as well as the left and right eigenvectors of $M$ are real and obey the relations $M \vec v_i=\lambda_i \vec v_i$, $\vec u_i M=\lambda_i \vec u_i$ and $\vec u_i^T=\Pi \vec v_i$.
Consider the spectral decomposition of  $M$:
\begin{equation}
M =\mathbf{1}  \pi + \lambda_2 \vec v_2 \vec u_2 + \lambda_3 \vec v_3 \vec u_3 +\ldots \quad
\end{equation}
and assume that the second largest eigenvalue $\lambda_2$ is not dominated by the smallest (i.e., $\lambda_2 > |\lambda_N|$), which is the case if the graph is not `almost bipartite'.
Then asymptotically we have:
\begin{equation*}
\lim_{t \to \infty} r_t(H) \sim  \lambda_2^t  \, \text{trace } H^T \vec v_2 \vec u_2 H.                                                                                         \end{equation*}
It is not difficult to see~\cite{Delvenne2010} that this leading term is maximised for the partition $H$ that classifies the nodes into two communities according to the sign of the entries of $\vec v_2$ (or, equivalently, $\vec u_2$).
This partition is called the normalised Fiedler partition~\cite{Shi2000}, a variant of the classic Fiedler partition proposed as a heuristic in spectral clustering~\cite{Fiedler1973,Fiedler1975}.

Therefore, spectral clustering, usually presented as a heuristic for the solution of combinatorial optimisation problems (see e.g. \cite{Shi2000}) emerges here as the exact solution of the large time scale stability optimisation.
It is worth noting that for the continuous-time case we can obtain similar results---here $P(t)$ behaves asymptotically as $e^{(\lambda_2-1)t} \vec v_2 \vec u_2$ leading to the same optimal partition.

\subsection{The meaning of the linearisation of stability at short times}
\label{sec:free_energy}

We now show that several heuristics for community detection that have been introduced from a statistical physics perspective appear from the linearisation of stability at small times.

Consider the continuous version of stability in the limit of small times  $t \to 0$ and expand to first order. This leads to the \textit{ linearised stability}~\cite{Delvenne2010,Lambiotte2009}:
\begin{equation}
\label{Eq:rlin1}
r_\text{lin}(t,H) = r(0,H) + t  \left. \frac{d r(t,H)}{dt}\right |_{t=0}= r(0,H) - t \,\, \text{trace}\left[H^T \frac{L}{2m} H\right].
\end{equation}

\subsubsection{Linearised stability as a type of free energy}
It then follows from~\eqref{eq:gen_cut} that
\begin{equation} \label{Eq:rlin}
 r_\text{lin}(t,H) = \text{ Diversity} - t \, \text{Cut}.
\end{equation}
This linearised stability can also be seen as a linear interpolation for the discrete-time stability between time zero and time one: $ r_\text{lin}(t,H) = (1-t)  r_0(H) + t \, r_1(H)$.
Incidently, this highlights a further connection between the discrete-time and the continuous-time stabilities, since the former can be seen as a linear approximation of the latter. We emphasize however that linearised stability is meaningful for all times, whether smaller or larger than one. Indeed, renormalising this expression as~\cite{Delvenne2010}:
\begin{equation}
 -\dfrac{r_\text{lin} (t,H)}{t} = \text{Cut} -\dfrac{1}{t}\text{ Diversity}
\end{equation}
makes it  clear that the linearised stability can be viewed as a free energy to be minimised: the diversity index acts as a type of entropy measure;  the Cut acts as an energy cost (as an analogy, think of the edges as bonds to be broken, whose energy is the weight of the edge);  and the inverse Markov time plays the role of temperature.

\subsubsection{Linearised stability is equivalent to the Reichardt \& Bornholdt Potts Model heuristic}
Reichardt \& Bornholdt (RB) proposed a Potts model heuristic for multiscale community detection based on the
minimisation of the Potts Hamiltonian  \cite{Reichardt2004,Reichardt2006}.
\begin{equation}
 \mathcal H_\gamma^\text{RB} = -2m \text{ trace }\left[  H^T\left(\dfrac{A}{2m} -\gamma\pi^T\pi\right)H\right] ,
\end{equation}
where $\gamma$ is a tunable resolution parameter.
From above, we observe that:
\begin{equation}
r_\text{lin}(t,H) = (1-t)-t\left( \frac{1}{2m} \mathcal H_{1/t}^\text{RB}\right).
\end{equation}
Hence maximizing the linearised stability is equivalent to minimising the RB Potts Hamiltonian
$\mathcal H_\gamma^\text{RB}$ with $\gamma = 1/t$.
Having established this connection enables us to provide an explicit interpretation of the `resolution parameter' $\gamma$ in terms of the Markov time $t$ and also to interpret the RB Potts model heuristic as a free energy, corresponding to a trade-off between entropy and cut~\cite{Delvenne2010}). Note that this view differs slightly from RB's initial interpretation in terms of energy only.

\subsubsection{Linearised stability with combinatorial Laplacian dynamics is related to the Constant Potts Model}

It is worth mentioning briefly that several other recent heuristics also correspond to the linearised stability obtained under a different dynamical evolution, namely when the dynamics is governed by the \textit{combinatorial} Laplacian~\cite{Lambiotte2009}:
\begin{equation}
 \mathbf{\dot p} =
 - \mathbf{p}\dfrac{L}{\langle d \rangle},
\end{equation}
where $\langle d \rangle$ is the average strength of the nodes in the graph.

It is then easy to show that maximising the linearised stability $r_\text{lin}^\text{CL}(t,H)$, which follows seamlessly from this dynamics, is equivalent to minimising the Hamiltonian of the so-called constant Potts model (CPM) introduced by Traag \textit{et al.} \cite{Traag2011}:
\begin{equation}
 \mathcal H_\mu^\text{CPM} = -2m \text{ trace }\left[  H^T\left(\dfrac{A}{2m} -\dfrac{\mu}{2m}\right)H\right],
\end{equation}
where $\mu$ is a tunable `resolution parameter.'
In fact,
\begin{equation}
 r_\text{lin}^\text{CL}(t,H) = (1-t) - t \left( \frac{1}{2m} \mathcal H_{2m/t N^2 }^\text{CPM} \right),
\end{equation}
and the optimization of the combinatorial linearised stability is equivalent to minimising the CPM with resolution parameter $\mu = 2m/N^2 t$.

A similar reasoning can be applied to the Potts heuristic proposed by Nussinov and coworkers~\cite{Ronhovde2009,Ronhovde2010}.

\section{Computational aspects of the stability framework}
\label{sec:optimisation}

\subsection{Optimisation of the stability measure}
Stability defines a quality measure on partitions and one can use in principle any graph clustering algorithm (even if based on different principles) and assess and rank the quality of the found partitions \textit{a posteriori} with our stability measure.
Some well-known algorithms that we have used in this way include the methods proposed by Shi and Malik \cite{Shi2000} and Kannan, Vempala and Vetta \cite{Kannan2000}, which operate divisively to obtain finer and finer partitions via recursive spectral bi-partitioning.
The partitions thus obtained can then be evaluated at different time scales according to stability: as outlined above, for short time scales we expect the finer partitions to be more relevant, whereas for larger time scales coarser partitions will dominate.
Using the spectral clustering methods in such a way turns them effectively into a heuristic for optimizing stability, which performs well in many cases \cite{Delvenne2010}.
See Section~\ref{sec:citation} for an example of this type of analysis.

Stability,  whether linearised, continuous-time or discrete-time, can also be considered an objective function to be optimised, i.e., for each time we have to find the partition $\hat H$ with maximal stability
\begin{equation}
 \hat H = \arg \max_H r(t,H).
\end{equation}
However, as is the case for most non-trivial problems in clustering and graph partitioning, including modularity optimisation \cite{Brandes2008}, the optimisation of stability over the space of all possible partitions is an NP-hard problem.
Therefore, except for asymptotic times $t\rightarrow 0$ and $t\rightarrow \infty$,  we must employ heuristics to optimise stability and these come with no guarantee on the obtained results.

A helpful feature of stability is the fact that it can be  written as the modularity of a time-dependent network evolving under the Markov process~\cite{Lambiotte2009}.
Therefore all the computational heuristics that have been developed for modularity optimisation can be naturally applied to stability optimisation too.
These include not only divisive spectral methods~\cite{Newman2006a} but also agglomerative methods, such as the efficient Louvain method~\cite{Blondel2008} based on a fast greedy optimisation that can deal with very large graphs.
Examples of the application of the Louvain algorithm to stability optimisation are shown in Sections~\ref{sec:grid}~and~\ref{sec:AS}. Other methods have been devised specifically 
for stability optimisation over a time interval~\cite{LeMartelot2012}.

One may as well combine different methods to optimise stability, at the expense of greater computational cost.  For instance, one can improve a Shi-Malik spectral partition with a local Kernighan-Lin scheme~\cite{kernighan1970efficient}.
A wide range of heuristic stability optimisation procedures can therefore be devised.

As is the case with hard non-convex optimisation problems, it is important to assess whether these different optimisation methods yield consistent results, i.e., to check that the value of stability obtained by the use of different optimisation algorithms is robust.
Our numerical experiments have shown that this is indeed the case specifically when the partitions found are most relevant~\cite{Delvenne2010,Delmotte2011,Schaub2012}.

\subsection{Assessing the robustness of a partition}
Related to the question of the robustness of the optimisation is the question of the robustness of the partitions found, which we use as an indicator of the relevance of the partitions found.
One simple mechanism to detect the robustness of a partition is already built into the stability measure via the Markov time: a robust partition should ideally be persistent over an extended range of time-scales.
When the graph has a strong community structure, this can be observed as plateaux extending over Markov time intervals during which each corresponding partition is found.

As a second indicator for the robustness of the partitions, we evaluate the dispersion induced in the optimised partitions by a randomisation of the optimisation algorithm.
In particular, for each Markov time we compute an ensemble of randomised Louvain optimisations of stability (started from a random initial node ordering) and check how different the optimised partitions are, as established by the variation of information \cite{Meila2007}.
Let $p(\mathcal C_\alpha)$ be the relative frequency of finding a node in community $\mathcal C$ in partition $\mathcal P^\alpha$, i.e., $p(\mathcal C_\alpha) = n_{\mathcal C_\alpha}/N$, where $n_{\mathcal C_\alpha}$ is the number of nodes in community $\mathcal C_\alpha$.
The variation of information between two partitions $\mathcal P^\alpha$ and $\mathcal P^\beta$ is defined as~\cite{Meila2007}:
\begin{equation}
\label{eq:VI}
 \text{VI}(\mathcal P^\alpha, \mathcal P^\beta) = 2 H(\mathcal P^\alpha, \mathcal P^\beta) - H(\mathcal P^\alpha) - H(\mathcal P^\beta)\\
\end{equation}
where $H(\mathcal P) = -\sum_{\mathcal C} p(\mathcal C) \log p(\mathcal C)$ is the Shannon entropy and $H(\mathcal P^\alpha, \mathcal P^\beta)$ is the Shannon entropy of the joint probability.
Here we use a normalised variant of the variation of information obtained by dividing the quantity in Eq.~\eqref{eq:VI} by its maximum, $\log N$.
A low variation of information indicates optimised partitions that are very similar to each other, indicating that the partition is robust to changes in the starting point of the optimisation.
Borrowing freely some terminology from dynamical systems, robust partitions can thus be considered to possess an attractor with a large basin of attraction for the optimisation process.

Finally, one can also consider the robustness of a partition to small changes in the underlying graph either through random perturbations~\cite{Lambiotte2010} or through the creation of surrogates for specific applications~\cite{Delmotte2011}.
All these measures of robustness are used to select the relevant partitions across Markov times.

\section{Examples of community detection with the stability framework}
In this section, we showcase a few applications of the stability framework.
Apart from the examples discussed below, this approach has already been applied to a variety of fields, such as image segmentation, social networks and computational biology, among others~\cite{Delmotte2011, meunier2010modular, Schaub2012, Delvenne2010}.

\subsection{A first example: community structure in a collaboration network}
\label{sec:citation}
\begin{figure}
\begin{center}
\includegraphics{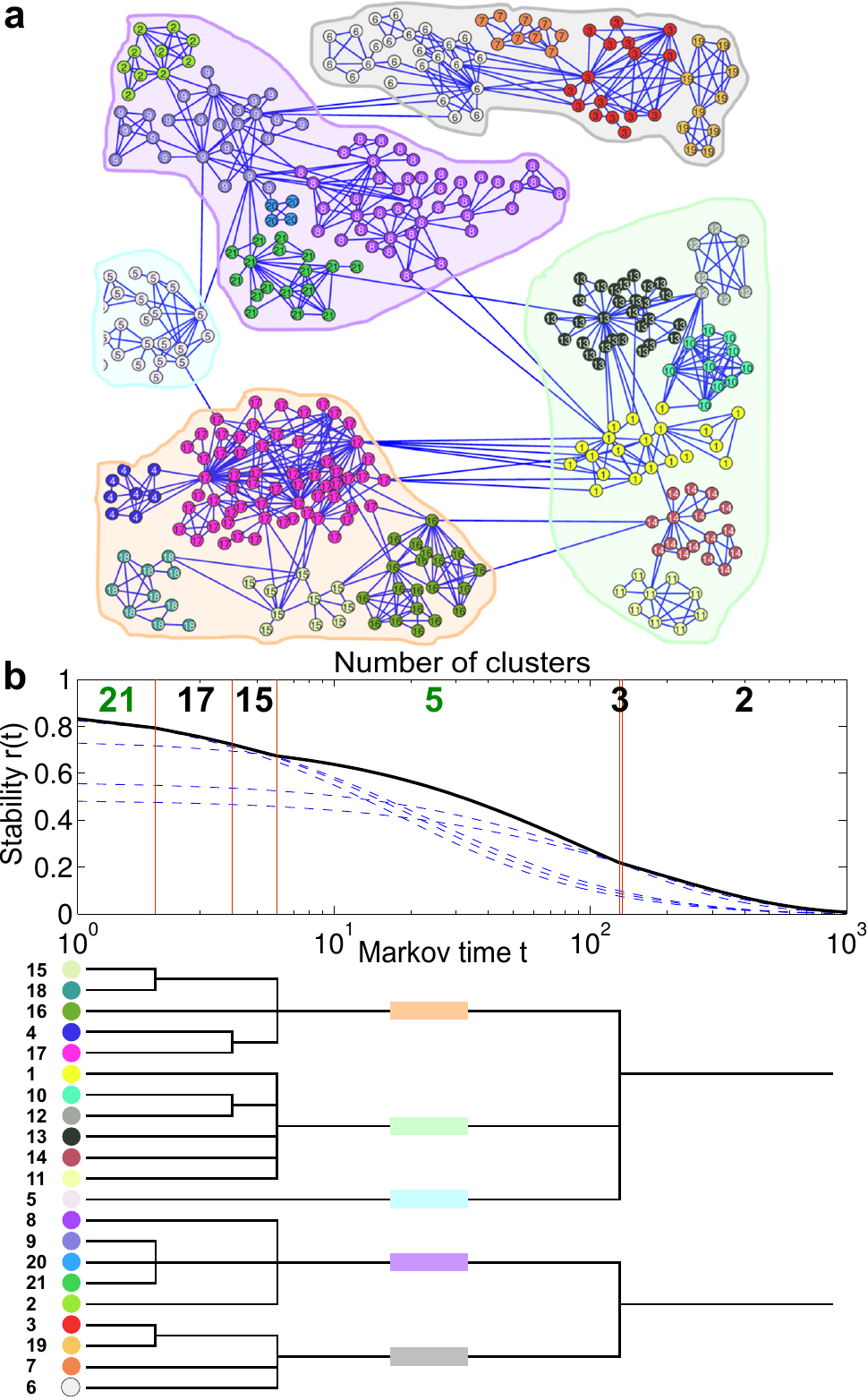}
\end{center}
\caption{\textbf{A network of scientific co-authorship~\cite{Newman06}}. \textbf{a} The vertices correspond to 379 researchers in network science, indexed by the 21-community partition obtained by maximizing the time-one discrete-time stability (i.e., modularity).  \textbf{b} We estimate the optimal partition for every time with a divisive KVV algorithm and exhibit the optimal stability curve as a function of time (top)  and the corresponding dendrogram of the hierarchy of partitions (bottom). The dendrogram is not a binary tree (as compared with most of the hierarchical clustering methods) and has relatively few levels. Except for the two-way partition, the decomposition into five communities (represented by the coloured shaded areas in \textbf{a}) stands out by its longevity. \cite{Delvenne2010}}
\label{FigNetSci}
\end{figure}

Community detection methods are commonly used for social networks, with the aim to identify sets of strongly interdependent people so as to shed light on the global structure of the social group.
As an example of this type of application, we analyse in Fig. \ref{FigNetSci} a graph of co-authorships between 379 researchers in network science (see \cite{Newman2006a}).
Figure~\ref{FigNetSci}b shows the hierarchy of partitions associated with the maximisation of discrete-time stability over all partitions obtained by using the Kannan, Vempala and Vetta (KVV) algorithm~\cite{Kannan2000} across all Markov times.
The KVV method is a conductance-based spectral divisive algorithm that splits the network in two parts repeatedly (in a kind of bisection scheme) producing a set of partitions that are then ranked according to stability at each time.
This procedure results in a streamlined \textit{non binary} dendrogram of optimal partitions, which differs from the notably more complicated binary trees produced by most hierarchical methods using different local heuristics.
The communities shown have been selected according to their long persistence in time.
Note in particular the very long-lived partition into 5 communities, which constitute well-formed thematic and national groups of authors, as a close look at the identity of the nodes reveals \cite{Delvenne2010}.

By construction, the particular optimisation algorithm used here as a heuristic for this NP-hard problem outputs a perfect hierarchy, where partitions are embedded into one another.
However, there are many instances where the actual partitions at different time scales are not hierarchical, i.e., they are not necessarily exact progressive refinements of each other.
As shown below, optimisation of stability with other algorithms can compute a non-hierarchical, yet multiscale, community structure across time scales.
One can then find if the sequence of partitions is close to being hierarchical or not~\cite{Lambiotte2009}.

\subsection{Non clique-like communities and the field-of-view limit}
\label{sec:grid}
\subsubsection{Stability as a multi-step method \textit{vs.} single-step methods}

In contrast to stability, most commonly used community detection algorithms can be reformulated as single-step methods from our dynamical viewpoint, i.e., these methods only take into account nearest neighbours in the graph  with no paths of length greater than one considered.
For modularity and derived measures (like the RB Potts model) this is obvious from our discussion in Section \ref{sec:relations}.
It can also be shown that the popular Map equation method \cite{Rosvall2008} is effectively based on single-step, block-averaged transitions of a random walker~\cite{Schaub2012, Schaub2011a}.
However, by using one-step measures, such methods effectively introduce an upper-scale into the detection of communities.
Metaphorically, these algorithms suffer from a limited ``field-of-view'' (see Ref.s \cite{Schaub2012,Schaub2011a} for an in-depth discussion) that precludes them from detecting communities with large internal distances.
In contrast, the stability framework has no preferred scale (or number of steps) \textit{a priori}, i.e. the scanning across all scales imposed by the dynamics is intrinsic and an essential feature of the algorithm.
Therefore, if there are one or more scales, they can be found through the dynamic sweeping.

\subsubsection{Advantages of a multi-step approach: detection of non clique-like communities}

If one were to consider the simplest notion of community structure, one would probably think of a graph
in which clique-like graphs with strong connections within them are weakly connected to each other in an 'all-to-all' manner.
The community structure thus emerges from the different intra- and inter-community weights leading to a  stochastic ``clique-of-cliques''.
The corresponding notion of community as a homogeneously connected dense substructure homogeneously embedded in the whole graph underlies most of the popular benchmarks for community detection~\cite{Lancichinetti2008,Lancichinetti2009a,Newman2004,Danon2006,Karrer2011}.
Although such notion of community structure is a good description in several application areas (e.g. networks constructed from correlation measurements or in some instances of social groupings), there exists a wide range of real networks that do not display such an ``all-to-all'' connection pattern.

Prominent examples of networks without clique-like connection patterns are geographically embedded networks, such as  sensor-networks, power grids, river-networks, road- and train-networks, as well as other transport, supply or distribution networks.
More generally, networks underpinned by intrinsic constraints dictated by geometry or other cost functions (e.g., higher-dimensional grid- or lattice-like structures originating from physical and biological systems) will not in general display homogeneous, block-like structures in their connectivity patterns.
Yet such networks may still have a pronounced community structure in its general sense: a \textit{non clique-like} community may in this case be thought of as a group of nodes which possess a stronger direct or \textit{indirect} influence on each other than on nodes outside their community, e.g., if entities are coupled via a chain of local interactions~\cite{Schaub2012}.
In these instances, one-step methods, such as modularity and the Map method, will fail to identify such non clique-like communities due to their large effective diameter, which puts them beyond the field-of-view of standard single-step methods.
It is important to remark that in the case of clique-like community structure, stability still recovers the results of standard methods, which can be seen as a particular case of our framework.

\begin{figure}[p!]
 \centering
 \includegraphics{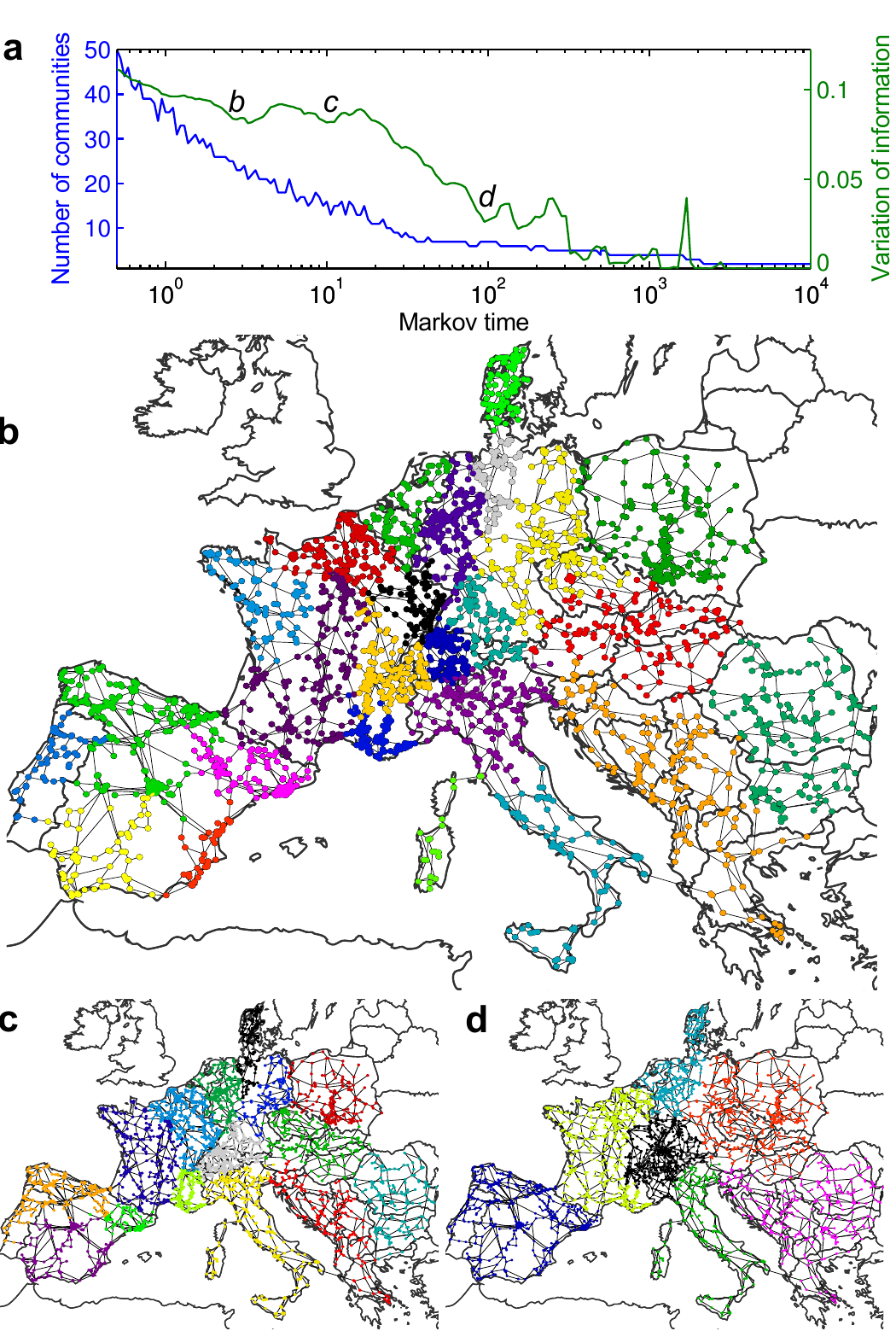} 
\caption{\textbf{Community structure analysis of the European power grid with stability}. \textbf{a} Stability analysis on the European power grid: number of communities found (blue) \textit{vs.} Markov time. A relatively low average variation of information distance \cite{Meila2007} (green) for a given Markov time indicates a robust partition. \textbf{b-d} The robust partitions for different Markov times seem consistent with known structures in the power grid. Those partitions correspond to Markov times: \textbf{b} $t = 2.63$ (25 communities), \textbf{c}  $t = 11.76$ (15 communities) and \textbf{d} $t=94.79$ (7 communities). Analysis of the same graph with popular methods, such as modularity and the Map equation, lead to severe overpartitioning, due to the intrinsic one-step nature of these methods~\cite{Schaub2012}.}
 \label{fig:1}
\end{figure}

\subsubsection{Analysis of the graph of the European Power Grid}
The analysis of the European Power Grid, a network in which non clique-like community structure plays a significant role, is used here to exemplify the advantage of using an intrinsic multi-step method such as stability~\cite{Schaub2012}.
This network\footnote{Dataset taken from \url{http://www.termoenergetica.upc.edu/marti/index.htm}} is based on data from the Union for the Coordination of Transmission Energy (UCTE) and has been analysed previously for robustness to targeted attacks~\cite{Rosas-Casals2007,Sole2008}.
It consists of 2783 nodes corresponding to generators and substations with 3762 (unweighted) links corresponding to high voltage (110 kV - 400 kV) transmission lines  (see also~\cite{Rosas-Casals2007,Sole2008}).
Geography and engineering costs impose some extrinsic local constraints on the Power Grid and hence its connection patterns are far from a homogeneous all-to-all type.
Hence we expect the relevant community structure to be of a non clique-like type.
The dynamic sweeping  of the stability framework over different Markov times is shown in Figure~\ref{fig:1}.
In this example, the analysis has been performed with the continuous time-variant of stability, optimised via the Louvain algorithm adapted for stability.
The Markov times highlighted in Figure~\ref{fig:1} have been selected according to the relative decrease in variation of information, which indicates a more robust partitioning.

The analysis reveals a multiscale community structure within the Power Grid, with distinct relevant partitions at various Markov times corresponding to subregions which, interestingly, reflect meaningful historical and commercial features of the power grid network.
For long Markov times (Fig.~\ref{fig:1}d), the coarse-grained subregions of the network can be identified mostly with nations or supra-national structures, which may be associated with big historical monopolies, while concurrently the German power grid remains split between four large companies (with one covering the eastern part of Germany)\footnote{See \url{http://de.wikipedia.org/wiki/Stromnetz\#Netzbetreiber} }.
For shorter times (Fig.~\ref{fig:1}b-c), communities on a sub-national scale are obtained. These are associated with regional operators, e.g., the French Power Grid is divided into several regions which overlap well with the communities found in our analysis\footnote{For a map of the French regional electrical companies see \url{http://www.rte-france.com/fr/nous-connaitre/qui-sommes-nous/organisation-et-gouvernance/le-siege-et-les-unites-regionales}}. Similar results hold for Spain, Italy and Switzerland. For more details see Ref.~\cite{Schaub2012}.

\subsection{Detecting communities of links in the Internet Autonomous Systems network}
\label{sec:AS}
\begin{figure}[p!]
 \centering
 \includegraphics{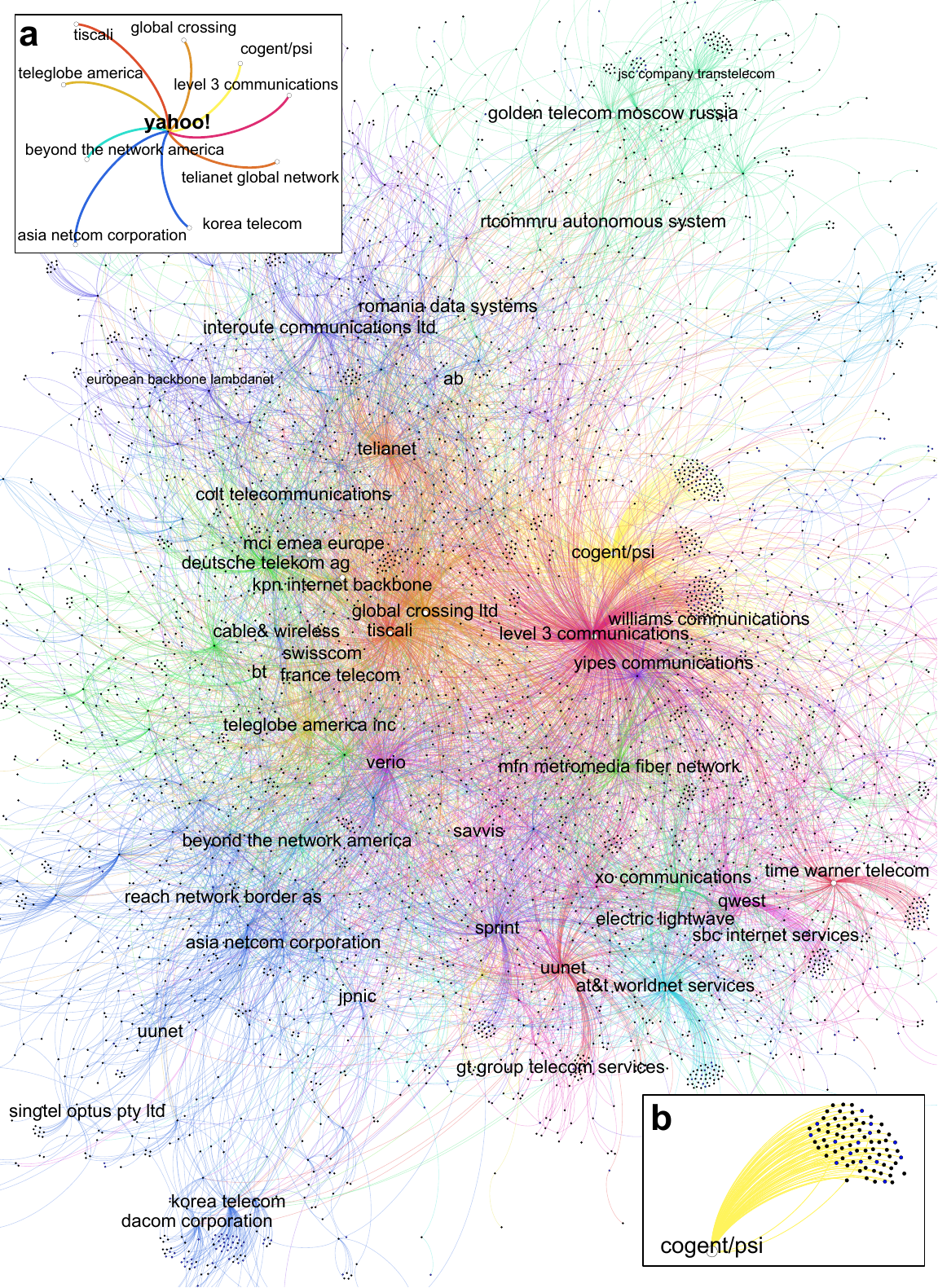}
 \caption{\textbf{Community analysis of an Internet AS network.} Community structure in the graph of edges of an AS network with 3951 nodes and 9870 (unweighted) links.  The links are grouped into different communities by optimising the continuous-time stability of the \textit{line graph} of this network. For Markov time $t= 0.52$, we find 31 communities for the edges (indicated by different colours). The nodes of the graph are coloured according to their type: t1 nodes in white, t2 in black, edu in blue. All 43 large ISPs (t1 nodes) in our dataset are labelled explicitly. Insets: \textbf{a} A small ISP node (yahoo!) is connected to several large ISP neighbours (t1) through edges that belong to different communities. \textbf{b} Example of a stub AS connected to the t1 node ``cogent/psi'' where all edges belong to the same community.}
 \label{fig:AS}
\end{figure}

As a final example, we look at another classical network system, namely the Internet at the level of Autonomous Systems (AS). An autonomous system is a large group of connected routing prefixes implementing a clearly defined, consistent routing policy\footnote{see \url{http://en.wikipedia.org/wiki/Autonomous_system_(Internet)}}. As this network is only known approximately, we use measurements obtained by the Cooperative Association for Internet Data Analysis (CAIDA)~\cite{CAIDA}.
To construct the network, we select only direct links between two different and unique AS and we ignore all Multi-Origin AS (MOAS) edges (i.e., links where a prefix originates from multiple AS).
Furthermore, we limit ourselves to AS of large Internet Service Providers (ISPs), small ISPs and universities corresponding to the categories `t1', `t2', and `edu' from the Autonomous System Taxonomy Repository\footnote{\url{http://www.caida.org/data/active/as_taxonomy/}}. Note that, in general, there is no equivalence between an ISP and a AS: although large enough ISPs tend to form their own AS, an AS can also consist of several smaller companies or correspond to the network of a university. The analysed network consists of 3951 nodes and 9870 unweighted, undirected links.

To exemplify the use of our framework with a different objective, we performed the community analysis on the space of edges.
This can be achieved by analysing the \textit{line graph} of the network.
The line graph of a network $G$ is a graph whose nodes are the edges of $G$, with a link between two nodes if the corresponding edges in $G$ are adjacent.
The study of the line graph allows us to focus on the links in the network, which carry the actual information between the AS,  to find whether there is a structural organization in the edges, corresponding to topological ``traffic-communities'' in the global network structure.
Since community labels are assigned to edges, rather than to the AS nodes, this allows for multiple community memberships of an AS node~\cite{Evans2009}.
The results of the community analysis of the line graph using the Louvain algorithm to optimise the continuous-time stability are shown in Figure \ref{fig:AS}.

Our analysis reveals interesting structural features. The links to/from each of the large ISPs belong predominantly to one cluster. For smaller ISPs, a different picture emerges. Some small ISPs appear to be \textit{stub AS}, in that they display only a small number of connections (usually one or two) linking them to one large t1 AS.  The links from the large t1 AS to its stubs typically all belong to the same community (Figure \ref{fig:AS}b). 
On the other hand, we find a second class of small ISPs which one may call \textit{diversified AS}:  they have edges belonging to different communities, connecting them to multiple large ISPs, suggesting that they ``diversify'' their connectivity patterns so as to enhance the robustness of their connectivity to the Internet. An example of such a small ISP with a diversified connection pattern is the AS of ``yahoo!'' with nine connections to different large ISPs (see Figure \ref{fig:AS}a) all belonging to different communities.

Clearly, the significance of these results is limited by the data and the techniques used to infer the connections between the AS, as well as by the fact that we do not consider the actual data volume transferred over the links.
However, the fact that this simplified analysis reveals interesting groupings within the AS network indicates the potential for future work in this direction. In particular, it may be interesting to test to what extent community detection might be used for a classification of the AS, possibly in comparison to what has been inferred by more classical measures. This will be the objective of future work.

\section{Conclusions and future work}
In this chapter we have shown that adopting a dynamical perspective towards community detection through the introduction of the stability measure provides us with a unifying framework that enables us to gain deeper understanding of community detection.
In particular it was shown how different community detection methods can be seen as special cases of the introduced stability and thus be interpreted from a dynamical viewpoint as well.

It has been further illustrated with a few examples how the stability framework can be used in different application areas.
The framework provides a surprisingly versatile set of tools and can be easily extended and adopted for different application scenarios.
It is important to mention that the framework naturally extends to directed networks \cite{Lambiotte2009}, another important class of networks which arise in wide spectrum of applications, especially when there is a flow between different entities in the network.

A whole set of new tools can be derived by considering different types of dynamics, including different types of random walks taking place on the network \cite{Lambiotte2009}.
This feature of the stability is especially interesting if one has some pre-knowledge about the flow pattern or the dynamics taking place on the network under consideration.
By adapting the dynamics to prior knowledge, one can potentially tailor the stability measure to the specific problem at hand and establish the most relevant community structure as it pertains to the dynamical functionality of the network.
In particular, our measure has been adopted for the study of time-evolving networks \cite{MuchaEtAl10}, an area of current focus in network research.

\section{Acknowledgements}
J.-C. D. acknowledges support from the grant ``Actions de recherche concert\'{e}es--- Large Graphs
and Networks'' of the Communaut\'{e} Fran\c{c}aise de Belgique, the EULER project (Grant No.258307) part of the Future
Internet Research and Experimentation (FIRE) objective of the Seventh Framework Programme
(FP7), and from the Belgian Network DYSCO (Dynamical Systems, Control, and Optimization) funded by the Interuniversity Attraction Poles Programme initiated by the Belgian State Science Policy Office.  S.N.Y. and M.B. acknowledge funding from grant EP/I017267/1 from the EPSRC (Engineering and Physical Sciences Research Council) of the UK under the \textit{Mathematics Underpinning the Digital Economy} program and from the Office of Naval Research (ONR) of the US.

\bibliographystyle{spphys}
\bibliography{bookchapter}

\end{document}